# Evolutionary algorithm for prediction of the atomic structure of two-dimensional materials on substrates

*Arslan B. Mazitov[1,2] *, Artem R. Oganov[3,]**

[1] Moscow Institute of Physics and Technology, 9 Institutsky lane, Dolgoprudny 141700, Russia
[2] Dukhov Research Institute of Automatics (VNIIA), Moscow 127055, Russia
[3] Skolkovo Institute of Science and Technology, Skolkovo Innovation Center, 3 Nobel Street, Moscow 121205, Russia

Corresponding Authors

* Arslan B. Mazitov, e-mail: arslan.mazitov@phystech.edu

* Artem R. Oganov, e-mail: a.oganov@skoltech.ru

**Today the study of two-dimensional (2D) materials has become one of the key objectives of materials science. Unlike their three-dimensional counterparts, 2D materials can simultaneously demonstrate unique transport and mechanical properties due to their dimensionality and quantum size effect. In their fabrication and application, 2D materials are usually located on top of the substrate or combined into heterostructures, which makes their structures and properties strongly depend on the nature and quality of the environment. Here, we present a novel method for studying the atomic structures of two-dimensional materials and epitaxial thin films on arbitrary substrates. The method can predict successful stages of epitaxial growth and the regions of stability of each atomic configuration with experimental parameters of interest. We demonstrate the performance of our methodology in the prediction of the atomic structure of $MoS_2$ on $Al_2O_3$ (0001) substrate. The method is also applied to study the CVD growth of graphene and hexagonal boron nitride on Cu (111) substrates. In both cases, stable monolayer and multilayer structures were found. The stability of all the structures in terms of partial pressures of precursors and temperature of growth is predicted within the *ab initio* thermodynamics approach.**

The field of 2D materials was widely developed after the first experimental production of graphene on a silicon oxide substrate by Novoselov and Geim in 2004[1]. Weak van der Waals interaction between layers in heterostructures of 2D materials, electron confinement inside the layers, and high surface-to-volume ratio lead to remarkable changes in electronic and optical properties of the materials, as well as in their chemical and mechanical response[2–7]. Besides, a wide range of ways to tune properties using lateral and vertical heterostructures fabrication[8–10], chemical functionalization[4,11], strain[12,13], defect[14,15] and substrate engineering[5], makes 2D materials an ideal candidates for developing a new class of electronic devices. According to International Technology Roadmap for Semiconductors[16], the use of 2D materials and their heterostructures in the fabrication of a new generation of transistors can improve the technological process from ∼ 5 nm to 1.5 - 2 nm by 2030. Moreover, many promising applications in the fields of photonics[17], photovoltaics[11,18], valleytronics[19], energetics[20], and catalysis[21] have already been realized in practice.

Fabrication of 2D materials typically relies on physical (PVD) and chemical (CVD) vapor deposition techniques, molecular-beam (MBE) and atomic-layer (ALE) epitaxy, or direct mechanical exfoliation method[7,22]. In each of these approaches, the material is finally located on top of a substrate. This fact makes its properties strongly depend on the nature and quality of the substrate. Indeed, experimental measurements show up to 80 % reduction in graphene's thermal conductivity on α-SiO$_2$ compared to freestanding graphene due to enhanced phonon-phonon scattering rate[5]. Furthermore, the substrate's roughness can lead to significant bandgap modulations[15,23], or even induce semi-metal - metal transition[24]. A correct choice of the substrate can also reduce inhomogeneity in charge density distribution and smoothen 2D material's landscape, as shown on the example of graphene on h-BN and SiO$_2$[25].

Another direction in the study of two-dimensional materials is the search for new instances of this class. A unique feature of almost all existing two-dimensional materials is their covalent nature of bonding[26]. However, in a recent experimental work[27] the possibility of forming two-dimensional NaCl and Na$_2$Cl on graphene surface by condensation from an unsaturated salt solution was shown. Previously, using the evolutionary algorithm (EA) USPEX[28–30], and subsequent experiments on the synthesis[31], it was already shown that the Na−Cl system has unexpected stable compounds at high pressure (Na$_3$Cl, Na$_2$Cl, Na$_4$Cl$_3$, NaCl, NaCl$_3$). More recent theoretical and experimental studies have shown that the two-dimensional hexagonal phase of rock salt (h-NaCl) can be stabilized on diamond (111) surface due to strong binding between salt film and substrate[32]. Such a wide variety of possible compounds suggests that even the simplest material, brought to unusual conditions, can exhibit unexpected properties.

This work introduces a method for predicting the atomic structure of 2D materials and thin films on arbitrary substrates, which is based on evolutionary algorithm USPEX. The new method allows one to automatically explore the whole range of atomic configurations and chemical compositions for given elements and measure their relative stability in the substrate's presence. We also use the *ab-initio* thermodynamics approach[33] to construct phase diagrams of growth, connecting the regions of stability of each atomic configuration with experimental parameters of interest, such as partial pressures of precursors in CVD and temperature of growth. The reliability of our technique is firstly tested on the prediction of the atomic structure of MoS$_2$ on c-cut sapphire (0001). Finally, we apply the method to study the CVD growth process of graphene and hexagonal boron nitride on Cu (111) substrate, using methane and borazine as precursors, respectively.

## Results

**MoS2 / Al2O3 (0001)**. We performed the first test of our method on the prediction of the atomic structure of MoS$_2$ within fixed composition evolutionary search. Among all 2D transition metal dichalcogenides (2D TMDs), MoS$_2$ is the most investigated material with a wide range of applications in next-generation optoelectronic devices due to its unique properties[34–36]. The most common bulk 2H-MoS2 has a layered crystal structure with $P6_3/mmc$ space group, a weak van-der-Waals coupling between layers, and strong covalent intra-layer bonding. With a decrease in dimensionality, a number of exotic metastable phases become possible[37]. Nevertheless, the most stable 1H phase inherits a trigonal prismatic Mo-S coordination from bulk crystal and has unit cell parameters of $a = b \approx 3.2$ Å, $\gamma = 120°$ and a Mo-S bond length of $\approx 2.4$ Å[38–40].

MoS2 monolayer preparation is usually based on various vapor deposition methods on SiO$_2$/Si, and $c$-cut sapphire substrates[35], while the most commonly used technology is a sulfurization of molybdenum oxide. In this work, we chose a (0001)-oriented sapphire as a substrate because of its chemical stability and symmetry, as in MoS$_2$. Since the unit cell of sapphire in $a - b$ plane is about 1.5 times bigger than the unit cell of MoS$_2$, a sufficiently large reconstruction is required to minimize the strain induced by lattice mismatch. Thus a 2x2 supercell of (0001) surface of sapphire was used as a substrate. The fixed-composition calculation contained 9 atoms of molybdenum and 18 atoms of sulfur in the unit cell.

The most stable structure found in the calculation is presented in Figure 1. The evolutionary algorithm successfully found the structure of 1H-MoS$_2$ as a 3x3 supercell. Corresponding parameters of the 1x1 unit cell are $a = b = 3.24$ Å, and Mo-S bond length is equal to 2.43Å. These values are reasonably close to the experimental measurements and theoretical predictions for freestanding MoS$_2$ [38–40].

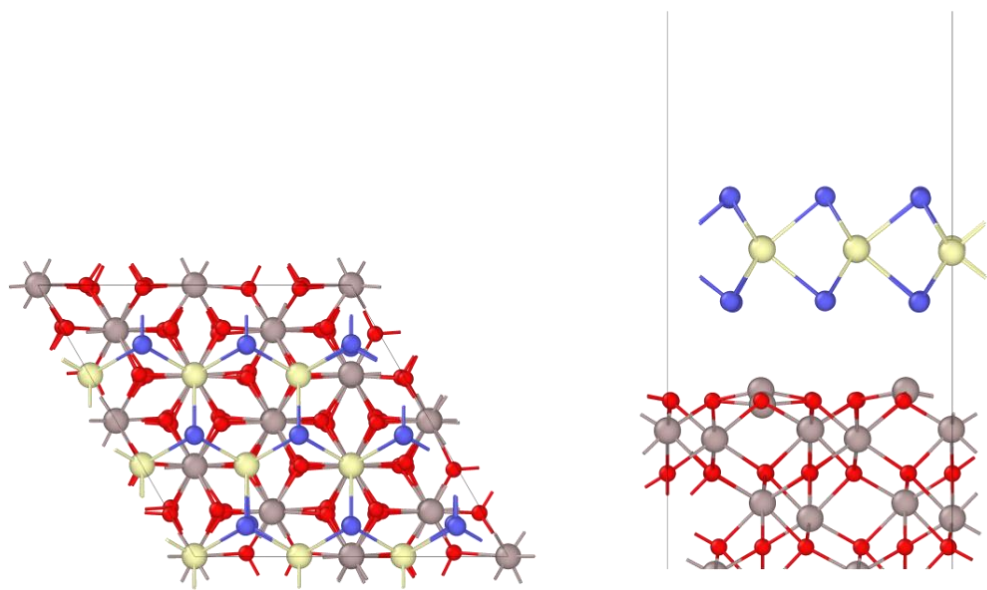

Figure 1. (Color online) Structure of MoS$_2$ on sapphire substrate: top (left) and side view (right). Molybdenum and sulfur atoms are represented by yellow and blue spheres, while aluminum and oxygen atoms are represented with gray and red spheres respectively.

**Graphene (C / Cu (111)).** A successful finding of the structure in fixed-composition search allows us to verify the algorithm in the prediction of the atomic structure of graphene with variable atomic density. Graphene has a recognizable honeycomb-like structure with $sp^2$ hybridized carbon atoms with bond length of 1.42 Å. Among all the preparation strategies, CVD has become the most promising due to low fabrication costs and high quality of graphene films[41,42]. Typically, graphene is grown on a variety of metal substrates, such as Cu, Ni, Ru, Ir, Pt, Co, Pd, and Re, using the decomposition of methane or ethylene as a source of carbon [43,44]: $CH_4 \rightarrow C + 2H_2$, $C_2H_4 \rightarrow 2C + 2H_2$. We performed our calculation using the Cu (111) substrate as the most popular among other metals due to extremely low carbon solubility[44]. The number of carbon atoms in the unit cell varied from 1 to 20, and the maximal surface area ratio of the unit cell and primitive cell of the substrate was equal to 4. Results of the calculation are presented in Figure 2 and Figure 3. The structure with a carbon density of 0.356 atoms/Å$^2$ is the structure of single-layered graphene. Final value of the corresponding 1x1 unit cell parameter is $a = 2.54$ Å with a bond length equal to 1.47 Å. These values differ from the equilibrium ones due to lattice mismatch, as was mentioned earlier.

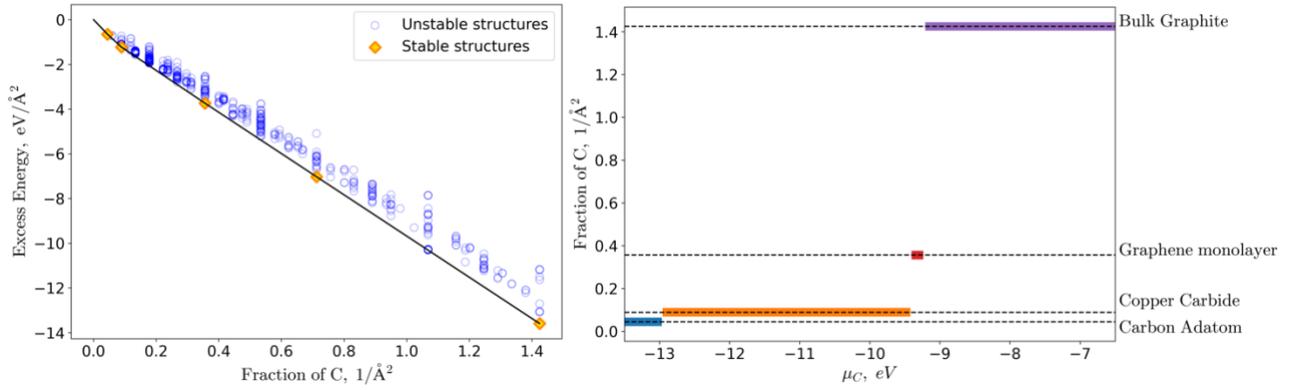

Figure 2. Convex hull (left) and phase diagram (right) of C / Cu (111) calculation. Stable structures on the convex hull are marked with orange diamonds, and unstable ones are marked with blue circles. Regions of stability in carbon chemical potential space are represented with color bars in the phase diagram.

Generally, the structures found can be interpreted as successive stages of carbon deposition on a copper surface. It starts from adsorption of single carbon atoms, as shown in Figure 3 (a,b). Then adatoms start to migrate due to surface diffusion and slightly dissolve in copper, which leads to the formation of intermediate stable clusters of copper carbide (Figure 3 (c,d)). It is interesting to note, that copper does not form bulk carbides, but here we find that surface carbide phases can (and should) exist. Further growth leads to the stabilization of single-layered and then quad-layered graphene, where the number of layers is controlled by the value of chemical potential of carbon (Figure 3 (e-h)). We note, that the structure of graphene quadlayer has the maximum possible number of atoms in the unit cell. Since there is no stable intermediate number of layers between graphene monolayer and quadlayer (or their regions of stability are extremely narrow), further increase of the number of atoms in the calculation will lead to stabilization of graphite with constantly growing height. Thus, we assume the quad-layered graphene to be the representation of bulk graphite.

It is worth noting the presence of diamond thin film structure in the calculation, which is metastable (Figure 4). Indeed, the same CVD techniques are widely used for diamond synthesis since the 1980s, including the use of copper substrates[45]. However, it requires a fine tuning of precursors composition and the temperature regime during the growth process, that cannot be described within our method.

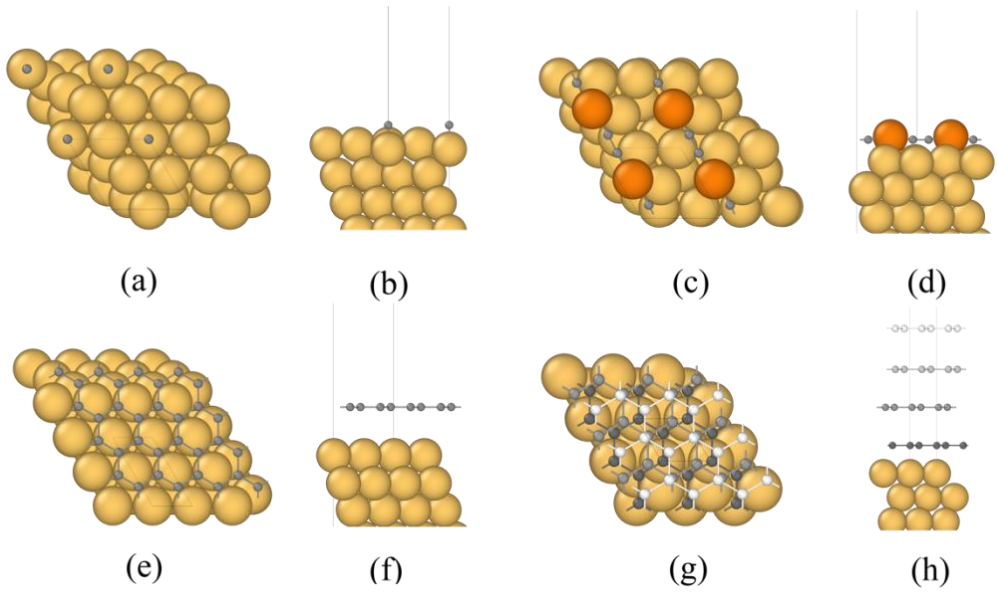

Figure 3. (Color online) Best structures of the C / Cu (111) calculation: single carbon adatom $C_{0.045}$ (a,b), Cu-$C_2$ nanowire $C_{0.089}$ (c,d), single-layer graphene $C_{0.356}$ (e,f) and bulk graphite $C_{1.425}$ (g,h). Atoms of copper are represented with large yellow spheres. Atoms of carbon are represented with small spheres and colored in greyscale depending on their height.

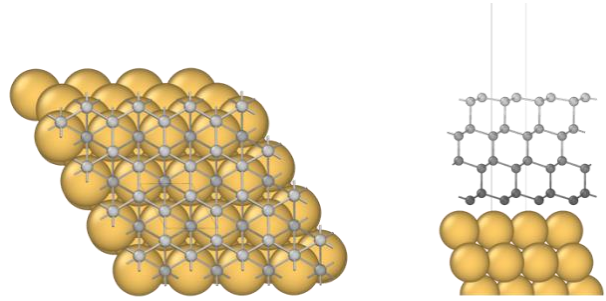

Figure 4. (Color online) Metastable diamond thin film: top (left) and side view (right). Color representation is similar to Figure 3.

The approach described in above allows one to calculate the phase diagram of CVD growth, which is represented in Figure 5. We used a reaction of decomposition of methane assuming thermodynamic equilibrium between reactants and products: $CH_4 = C + 2H_2$. This allows us to express the chemical potential of carbon through the chemical potentials of methane and hydrogen:

$$\mu_C(T, p_{CH_4}, p_{H_2}) = \mu_{CH_4}(T, p_{CH_4}) - 2\mu_{H_2}(T, p_{H_2}) \tag{1}$$

where $\mu_{CH_4}(T, p_{CH_4})$ and $\mu_{H_2}(T, p_{H_2})$ can be obtained using the formula (also see Methods section):

$$\mu_i(T, p_i) = \mu_i(T, p_{i,0}) + k_B T \ln\left(\frac{p_i}{p_{i,0}}\right) \tag{2}$$

The calculated values of the internal energy for methane and hydrogen molecules were -24.05 and -6.77 eV, respectively, while the values of zero-point corrections were equal to 1.18 and 0.27 eV.

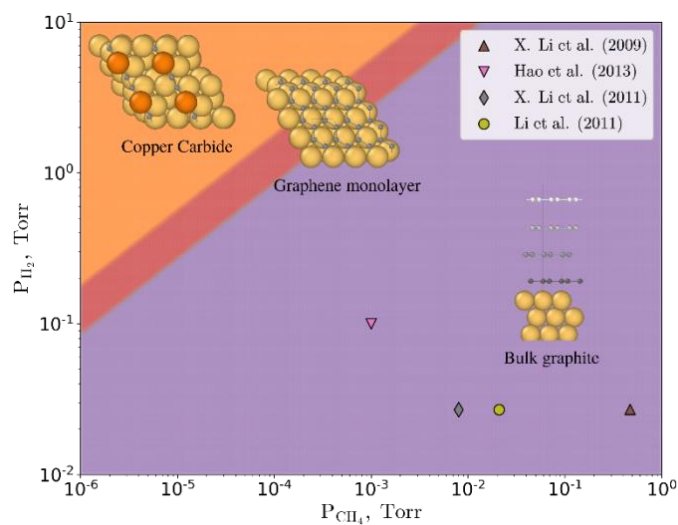

Figure 5. (Color online) Predicted phase diagram of graphene CVD growth on Cu (111) substrate at T = 1000 °C. Regions of stability of different structures are represented with colors. Experimental data on graphene monolayer growth conditions from works[46–48].

On the phase diagram, almost all experimental points are located far from the region of stability of graphene monolayer. This result indicates that the growth process is controlled by the kinetics of precursor decomposition and growth time. In other words, the thermodynamic equilibrium in reported CVD results[46–48] was not achieved. Indeed, the decomposition of methane at presented conditions ($P_{CH_4} = 10^{-2}$ Torr, $P_{H_2} = 10^{-2}$ Torr and $T = 1000$ °C) is energetically overdriven with $\Delta G_r$ of -1.74 eV (see Eq. S7 in Supplementary Information). In this way, if temperature and gas pressure remain constant, all the available methane decomposes into carbon layers and hydrogen molecules. More details on thermochemistry of decomposition process are available in Supplementary Information.

**h-BN (BN / Cu(111)).** Atomically thin hexagonal boron nitride is a notable 2D material widely researched for the development of perspective electronic devices, and the fabrication of other 2D materials [49–51]. It has very similar to graphene structure with a slightly longer bond of 1.44 Å. Similar to graphene, synthesis of h-BN is often performed by CVD. Originally, this process included the usage of gaseous boron-containing compounds in an ammonia atmosphere as precursors ($BF_3/NH_3$, $BCl_3/NH_3$, $B_2H_6/NH_3$)[52], though in recent works precursors with stoichiometric content of boron and nitrogen such as borazine ($B_3H_6N_3$, BZN) and ammonia borane ($BH_3NH_3$, AB) became more popular[53]. The latter are much more preferable since they do not require precise control of the gas mixture's composition and are also less toxic. The most popular substrates for CVD growth of boron nitride monolayers are metallic foils (Cu, Co, Ni, Pt), $SiO_2$/ Si substrates, and sapphire crystals[53,54]. Hence, we studied the growth of h-BN at the surface of Cu (111) substrate. We included up to 20 atoms of boron and nitrogen in the unit cell, and the maximal surface area ratio of the unit cell and primitive cell of the substrate was equal to 4.

Our results are summarized in and Figure 6 and Figure 7. Again, the structure of hexagonal boron nitride monolayer was successfully found in the evolutionary search. Since the shape of the unit cell is strictly defined by the substrate, h-BN structural parameters are similar to those of graphene: $a = 2.54$ Å and B-N bond length is equal to 1.48.

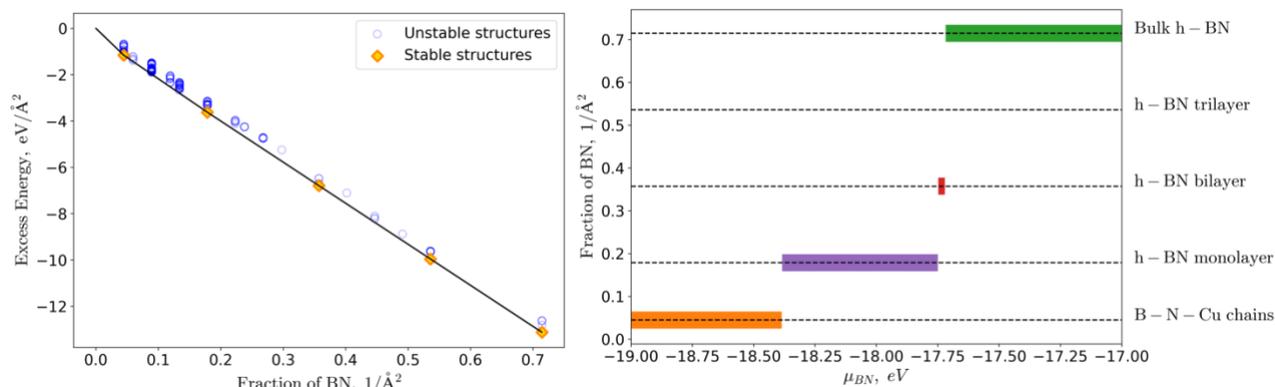

Figure 6. Convex hull (left) and phase diagram (right) of BN / Cu (111) calculation. Stable structures on convex hull are marked with orange diamonds, and unstable ones are marked with blue circles. Regions of stability in boron nitride chemical potential space are represented with color bars in the phase diagram.

Other found structures represent consequent steps of h-BN growth. In the beginning, the thermodynamically most preferable structure is a B-N-Cu chain, where the atom of copper is pulled out of the substrate. With an increase in $\mu_{BN}$, the hexagonal boron nitride monolayer structure can be stabilized. It is worth noting that regions of stability of bi- and tri-layer of h-BN are extremely narrow, while the monolayer structure is stable in a wide range of growth conditions. Further increase of chemical potential leads to the formation of bulk h-BN, which is represented here as a four-layer structure and corresponds to the maximum possible number of atoms in the calculation.

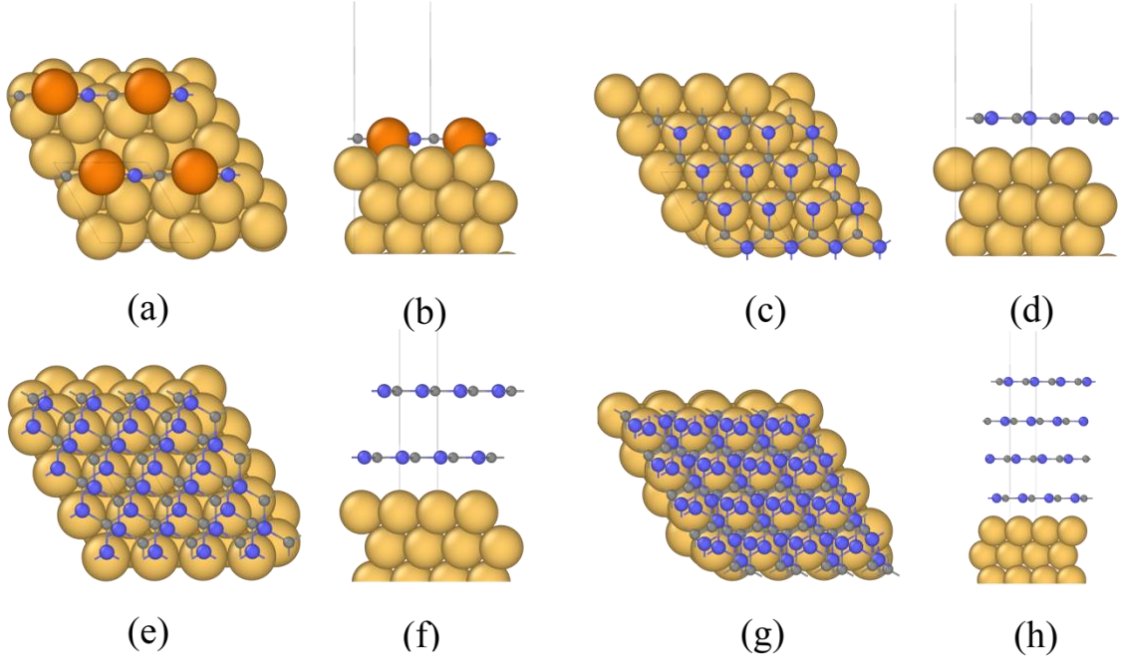

Figure 7. (Color online) Best structures of the BN / Cu (111) calculation: B-N-C nanowire $(BN)_{0.045}$ (a,b), hexagonal boron nitride monolayer $(BN)_{0.179}$ (c,d), h-BN bilayer $(BN)_{0.357}$ (e,f) and h-BN quadlayer $(BN)_{0.714}$ (g,h). Atoms of copper are represented with large yellow spheres. Boron and nitrogen atoms are represented by small gray and medium blue spheres, respectively.

In this work, we considered a reaction of dehydrogenation of borazine as the main source of boron nitride. This reaction is notable, because it also takes a place as a final stage of h-BN growth from ammonia borane, in addition to usage of borazine itself as a precursor. Thus, the two most commonly used methods of growth can be described within this approach. Detailed information on decomposition paths and thermochemistry of both AB and BZN can be found in Supplementary Information.

As it was done in the case of graphene, the dehydrogenation is assumed to be at thermodynamic equilibrium: $B_3H_6N_3 = 3BN + 3H_2$. It allows us to express the chemical potential of BN in the following way:

$$\mu_{BN}(T, p_{BZN}, p_{H_2}) = \frac{1}{3}\left(\mu_{BZN}(T, p_{BZN}) - 3\mu_{H_2}(T, p_{H_2})\right) \tag{3}$$

Corresponding calculated values of potential energy and ZPE for BZN are equal to -71.95 eV and 2.48 eV. We finally converted our data into phase diagram of h-BN growth on Cu (111) substrate at $T = 1000$ °C (Figure 8) and compared our results with experimental data on h-BN monolayer growth conditions from works[54–59].

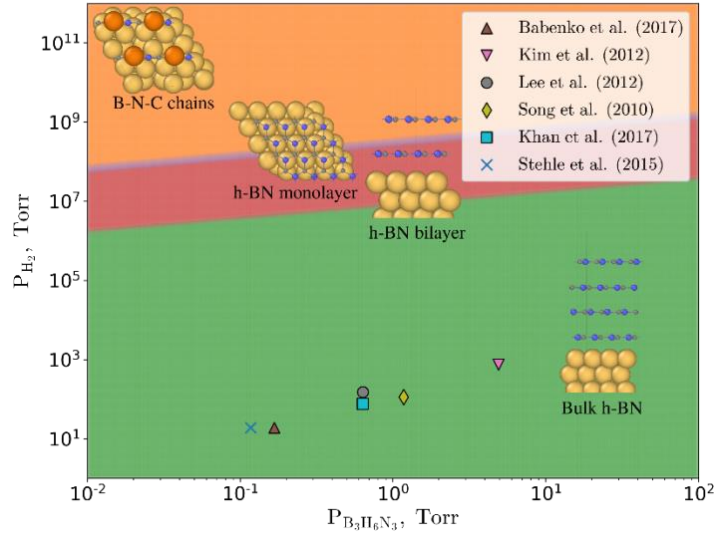

Figure 8. (Color online) Predicted phase diagram of h-BN CVD growth on Cu (111) substrate at T = 1000 °C. Regions of stability of different structures are represented with colors. Experimental data on h-BN monolayer growth conditions from works[54–59].

Similar to previous calculation, all the experimental data is concentrated far from the predicted region of stability of monolayer h-BN. This result is not surprising and can serve as another confirmation of non-equilibrium conditions in CVD growth. Gibbs free energy of decomposition of borazine (Eq. S9) at growth conditions ($P_{BZN} = 10^0$ Torr, $P_{H_2} = 10^2$ Torr and $T = 1000$ °C) is equal to -4.54 eV, i.e. growth in experimental conditions is thermodynamically overdriven, which supports our results once again (see Supplementary Figure 2).

## Discussion

Our results show that the CVD growth of such common 2D materials as graphene and hexagonal boron nitride is mostly determined by the kinetics of precursors decomposition and adsorption/desorption rates. In both cases, growth is thermodynamically favorable, according to Gibbs energy of decomposition of reactants. Our method, predicting thermodynamic phase diagrams, can still give qualitative description of the growth process for arbitrary combinations of deposited material and substrate. To sum up, in this paper, we presented a novel method of analysis of epitaxial thin film growth. The combination of evolutionary algorithm and accurate DFT calculations provides users with a powerful tool for exploring the set of possible atomic configurations during the growth process with variable stoichiometry and different coverage on arbitrary substrates. *Ab initio* thermodynamics approach allows to connect the results of the evolutionary search with experimental results of CVD growth at arbitrary conditions and precursors used through the construction of phase diagrams in the coordinates of interest. Our method was tested on the prediction of the crystal structure of 2D molybdenum diselenide on *c*-cut sapphire, where the most stable 1H-MoS$_2$ monolayer was successfully found. Thereafter we studied CVD growth of graphene from methane and hexagonal boron nitride from borazine on Cu (111) substrate. In both cases, the number of specific honeycomb-like layers were found. Direct comparison of growth conditions with constructed phase diagrams shows that the thermodynamic approach is still insufficient since the growth is mainly determined by kinetics and not by thermodynamics. Despite this fact, our method can be used to study the geometry of epitaxial structures and their phase transitions induced by interaction with the substrate, while the resulting thermodynamic phase diagrams can be used as a qualitative guide for synthesis. Epitaxial thin films are a fundamentally important and potentially unlimited class of atomic systems with unusual chemistry and unique properties, and the first step of their study can be done by predicting their structure with the presented technique.

## Methods

**Evolutionary algorithm.** Our methodology is based on the evolutionary algorithm USPEX[28–30], which has been previously used to predict the crystal structure of bulk materials, two-dimensional materials in a vacuum[60], reconstructions of surfaces[61] and nanoclusters[62]. Each structure is represented as a combination of three parts: substrate, surface, and vacuum (Supplementary Fig. 1). At the beginning of the calculation, the user is offered to define the chemical system to explore, set parameters of the evolutionary search, thresholds for the size of the film and vacuum layer, and provide the structure of the substrate. The algorithm then optimizes the structure of the film. The workflow is organized as follows. First, the initial structures (referred to as initial generation) in the film region are created using a topological random structure generator[63]. Each structure is then joined with the substrate and relaxed using external ab initio codes. Final structures are ranked by fitness function based on their energy. The following generations are produced by variation operators that use the previous generation's best structures as parents. In our method, we use four different variation operators to create offsprings from parent structures:

- Heredity (Crossover): Creates an offspring from two parents, slicing both parents along a randomly chosen axis and combining fragments in alternate order.

- Mutation: Creates an offspring from a single parent by shifting all atoms in the film along an eigenvector of one of the softest vibrational modes.

- Add/Delete: Creates an offspring from a single parent by adding an atom to the system or deleting an atom. Candidates for deletion and the location of new atomic sites are chosen according to coordination numbers of atoms in the parent structure.

- Transmutation: Creates an offspring from a single parent by permutation of chemical species of some atoms.

A certain number of structures in a generation are still produced randomly to diversify the population. After joining the substrate, offsprings with some fraction of the previous generation's best structures form a new generation. We also note that the primitive unit cell of the substrate might not be suitable for the film, especially if a significant lattice mismatch of bulk analog of the film and substrate takes place. This may be taken into account by selecting different basis vectors for the unit cell of the substrate. Hence, for each structure, we choose a basis as a random linear combination of the substrate's original basis vectors. The modulus of final basis vectors is restricted by setting the maximum ratio between areas of the unit cells of the film and the substrate. When a new generation is produced, structures are relaxed, and their energies are evaluated. This process is repeated cyclically until the best structures' set remains unchanged for a certain number of generations.

**Evaluation of stability of thin films.** The theoretical description of thin film stability is a key problem of this work. Usually, systems with variable composition and number of atoms are considered in the grand canonical ensemble, where the stability of structures is determined by composition-dependent free energy of formation. Hence it requires the knowledge of the chemical potential of components in the system:

$$G_{form}(T,p) = G_{tot}(T,p) - G_{ref}(T,p) - \sum_i n_i \mu_i(T,p) \qquad (4)$$

where $G_{tot}(T,p), G_{ref}(T,p)$ are Gibbs free energies of the whole system and of the reference system respectively, $n_i, \mu_i(T,p)$ are the number of excess atoms and chemical potential of component $i$. This concept involves the fact that the system is considered in thermodynamic equilibrium with a reservoir of components $n_i$, which defines the value of $\mu_i$. Generally, crystal growth process itself is a nonequilibrium process. However, the equilibrium thermodynamics approach remains relevant for most slow growth processes from the melt (or melt solutions). The growth from hot gas phases with sufficient density (not too far below ambient pressure of 1 bar, such as chemical vapor deposition (CVD) or physical vapor deposition (PVD)) is also expected to be described well within this approach[64]. We focus on these cases in further discussion.

We used so-called *fitness* function to indicate the relative stability of the structures with various compositions and sets of basis vectors during the evolutionary search. Algebraically it is equivalent to the energy of formation per unit area:

$$\gamma(T,p) = \frac{1}{A}\left(E_{tot} - E_{ref} - \sum_i n_i \mu_i(T,p)\right) \qquad (5)$$

where $A$ is an area of the unit cell. Here we approximate Gibbs free energies of structures during the evolutionary search by their internal energies to reduce computational complexity. Since the values of the chemical potentials $\mu_i$ are not known in advance, it is more convenient to define the excess energy of the structure, $\gamma_{exc}$, as $\mu_i$-independent term in (5): $\gamma_{exc} = \frac{1}{A}(E_{tot} - E_{sub})$ and represent all the structures in EA search as a set of points in excess energy - atomic density ($\gamma_{exc}, n_i/A$) space. Therefore, the set of stable structures forms a convex hull on such a diagram (Supplementary Fig. 2). For more details about this approach, the reader is referred to works[61,65,66].

The value of chemical potential geometrically corresponds to the slope of the section of the convex hull at given composition. Those structures that form the section can coexist in equilibrium at the corresponding value of $\mu_i$. The evolutionary algorithm then optimizes the convex hull in the same way as it was implemented in our previous works[28,60–62]. Taking partial derivatives of the convex hull with respect to atomic density, $\mu_i = \partial \gamma_{exc}/\partial(n_i/A)$, one can calculate the regions of stability of phases in chemical potentials space (Supplementary Fig. 3). This approach can be extended to the systems with an arbitrary number of components with a multi-dimensional convex hull, based on extended space of atomic densities.

**Construction of the growth phase diagrams**. The successful EA search can not only predict the set of stable structures but also give the regions of stability in chemical potentials space. This allows the *ab-initio* thermodynamics[33] to be used for calculating the growth phase diagrams. Modeling of the vapor phase growth is usually performed in the ideal gas approximation, where the dependence of chemical potential on temperature $T$ and partial pressure $p_i$ of gas phase can be expressed as follows.

$$\mu_i(T,p_i) = \mu_i(T,p_{i,0}) + k_B T \ln\left(\frac{p_i}{p_{i,0}}\right) \qquad (6)$$

where $E_{i(gas)}$ is a potential energy of an atom (or a molecule) from *ab-initio* calculation, $p_{i,0}$ is a reference pressure (usually $p_{i,0} = 1$ bar), $k_B$ is a Boltzmann constant. The value of $\mu_i(T, p_{i,0})$ can either be taken from NIST-JANAF thermochemical tables[67] in form

$$\mu_i(T, p_{i,0}) = E_{i(gas)} + E_{ZPE} + [H(T, p_{i,0}) - H(0, p_{i,0})] - T\left(S(T, p_{i,0}) - S(0, p_{i,0})\right) \quad (7)$$

or calculated within standard thermochemical procedures[68] (see also section 6.1 in Supplementary Information).

In the previous subsection, we neglected the temperature-dependent contribution in Eq. (5). However, the consistent calculation of formation energies is possible only if zero-point correction and vibrational entropy of the structures are also considered. Thus, for each stable structure (i.e., for structures on the convex hull) and substrate, we estimated this contribution in the quasi-harmonic approximation. Finally, the temperature-dependent energy of formation can be expressed as follows:

$$\gamma(T, p) = \frac{1}{A}\left(F_{tot}(T) - F_{ref}(T) - \sum_i n_i \mu_i(T, p)\right) \quad (8)$$

where $F(T) = E + E_{ZPE} - TS_{vib}$ is the Helmholtz free energy.

Substituting the values of $\mu_i(T, p_i)$ into Eq. (8), one can calculate the values of formation energies for each stable structure at each value of $T, p$. Thus, the regions of stability of structures can be determined by finding the lowest $\gamma(T, p)$ for each $T, p$. This information can be clearly demonstrated in the phase diagram of the growth process. An example of such a phase diagram is shown in Supplementary Fig. 4.

**Details of the ab-initio calculations.** The workflow of our algorithm typically requires hundreds (or even thousands) of structure relaxations. In this work, we performed relaxations using density functional theory implemented in the VASP[69,70] software package. Core electrons and their effect on valence electrons were described within projector augmented wave[70,71] potentials, and exchange-correlation interaction was expressed in the generalized gradient approximation with Perdew-Burke-Ernzerhof functional. Cutoff energy for plane waves basis set was 500 eV, and the first Brillouin zone was sampled by a $\Gamma$-centered grid with $2\pi \cdot 0.05$ Å$^{-1}$ resolution. Dipole correction along the normal vector to the film's surface was also utilized to cancel long-range interaction through periodic boundary conditions. DFT-D3[72] correction was included to take van der Waals interactions into account. Phonon energies and density of states were calculated using the PhonoPy[73] package.

**Details of the evolutionary search.** All the evolutionary search parameters in our work were rather similar and distinguished only in the chemical composition of the thin film under consideration. The first generation in each calculation was produced randomly, while the subsequent ones were generated by applying heredity (30 %), softmutation (20 %), transmutation (10 %) operators, and add/delete (10 %) operators. The remaining 30 % were produced randomly. The thicknesses of the vacuum and film region were set to 20 Å and 5 Å respectively. All the atoms of the substrate that were deeper than 3 Å from the surface were frozen during the relaxation.

# Acknowledgements

The authors are deeply grateful to Alexey V. Yanilkin for his help and useful discussions. A.B.M thanks Russian Science Foundation (grant No. 18-73-10135) for financial support. Work of A.R.O. is supported by Russian Science Foundation (grant 19-72-30043).

# Author contribution

A.B.M designed the extension of USPEX code, performed the calculations and wrote the original draft. A.R.O directed the project, provided theoretical support, contributed to the interpretation of the results, reviewed and edited the paper.

# Data availability

The data generated and/or analyzed during the current study are available from the corresponding authors on reasonable request.

# Code availability

Code and accompanying scripts developed in the current work are available upon request.

# Evolutionary algorithm for prediction of the atomic structure of two-dimensional materials on substrates

*Arslan B. Mazitov[1,2]*, Artem R. Oganov[3,]**

[1] Moscow Institute of Physics and Technology, 9 Institutsky lane, Dolgoprudny 141700, Russia
[2] Dukhov Research Institute of Automatics (VNIIA), Moscow 127055, Russia
[3] Skolkovo Institute of Science and Technology, Skolkovo Innovation Center, 3 Nobel Street, Moscow 121205, Russia

Corresponding Authors

* Arslan B. Mazitov, e-mail: *arslan.mazitov@phystech.edu*

* Artem R. Oganov, e-mail: a.oganov@skoltech.ru

# Supplementary Information

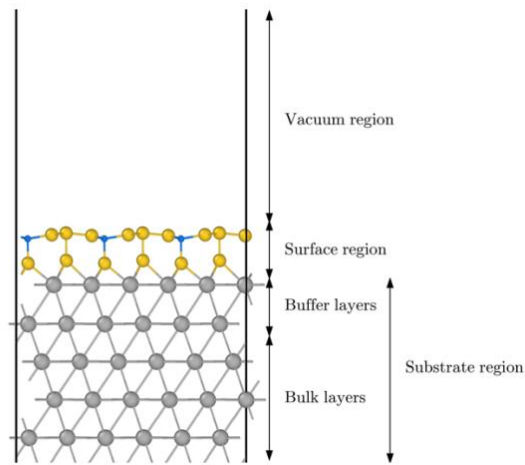

Supplementary Figure 1. Representation of epitaxial structures in our evolutionary algorithm. The substrate region is divided into two regions: buffer (which is released) and bulk (which is kept fixed). Evolutionary algorithm optimizes the structure in surface region. Vacuum layer is usually added to exclude the interaction of film and substrate through periodic boundary conditions.

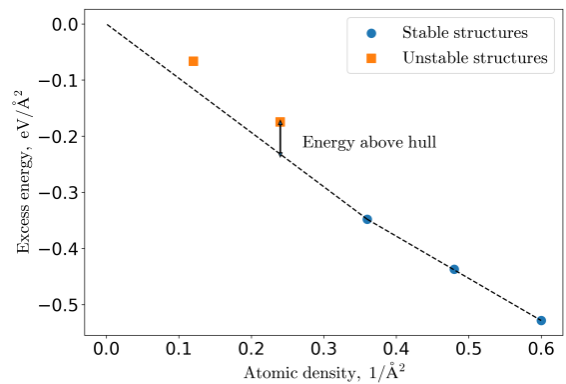

Supplementary Figure 2. (Color online) Convex hull diagram for a single-component film in $(\gamma_{exc}, n_i/A)$ space. Stable structures are marked with blue circles, and unstable ones are marked with orange squares. A convex hull of the diagram is drawn with a dashed line. The value of energy above the hull indicates the degree of instability of structures.

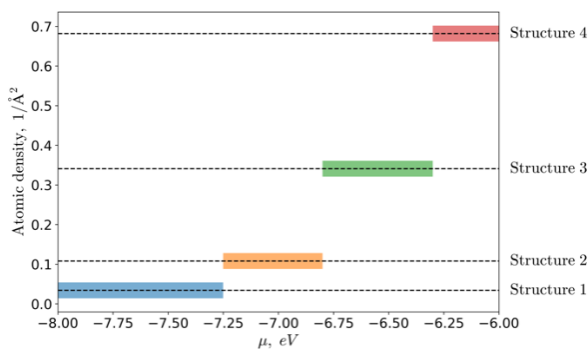

Supplementary Figure 3. Phase diagram in chemical potentials space. The stability region of each structure is represented by a color bar.

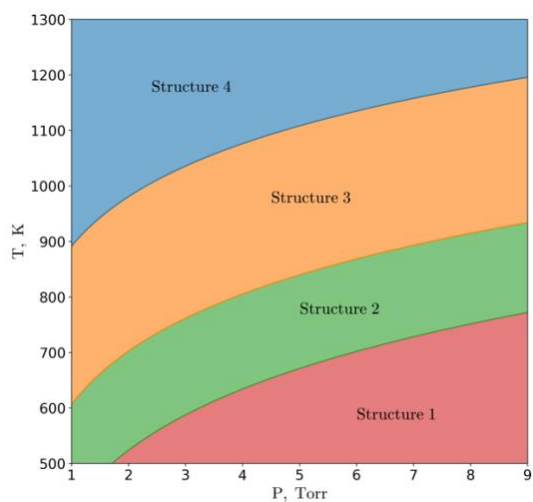

Supplementary Figure 4. Phase diagram of the epitaxial system in $(p, T)$ space. Colors represent regions of stability of different structures.

## Thermochemical details: theoretical background

All the gases in our work were considered in ideal-gas limit, assuming that separation on translational, rotational and vibrational degrees of freedom is valid. Thus, one can calculate the ideal-gas enthalpy as a sum of internal energy at 0 K and the integral over constant pressure heat capacity:

$$H(T) = E + E_{ZPE} + \int_0^T C_P \, dT \tag{S1}$$

where

$$C_P = k_B + C_{V,trans} + C_{V,rot} + C_{V,vib} \tag{S2}$$

is separated to the translational, rotational, and vibrational contributions. For a 3D gas $C_{V,trans}$ is equal to $3/2\, k_B$, while $C_{V,rot}$ is 0 to monoatomic species, $k_B$ for linear molecules and $3/2\, k_B$ for nonlinear molecules. The vibrational heat capacity in its integrated form can be represented as the sum over $3N-5$ (or $3N-6$) vibrational degrees of freedom for linear (or nonlinear) molecules:

$$\int_0^T C_{V,vib} \, dT = \sum_i \frac{\epsilon_i}{e^{\frac{\epsilon_i}{k_B T}} - 1} \tag{S3}$$

where $\epsilon_i = \hbar \omega_i$ are the energies of the corresponding phonons with the frequencies $\omega_i$.

The entropy of the ideal gas of molecules consists of corresponding translational, rotational and vibrational contributions as well:

$$S(T,P) = S_{trans}(T, P_0) + S_{rot}(T, P_0) + S_{vib}(T, P_0) - k_B \ln P/P_0 \tag{S4}$$

where

$$S_{trans}(T, P_0) = k_B \left( \frac{5}{2} + \ln \left[ \frac{\left( \frac{2\pi M k_B T}{h^2} \right)^{\frac{3}{2}} k_B T}{P_0} \right] \right)$$

$$S_{rot}(T) = \begin{cases} 0, & \text{for monoatomic gas} \\ k_B \left( 1 + \ln \frac{8\pi^2 I k_B T}{\sigma h^2} \right), & \text{for linear molecules} \\ k_B \left( \frac{3}{2} + \ln \frac{\sqrt{\pi I_A I_B I_C}}{\sigma} \left( \frac{8\pi^2 k_B T}{h^2} \right) \right), & \text{for nonlinear molecules} \end{cases} \tag{S5}$$

$$S_{vib}(T) = k_B \sum_i \left( \frac{\epsilon_i}{k_B T \left( e^{\frac{\epsilon_i}{k_B T}} - 1 \right)} - \ln \left( 1 - e^{-\frac{\epsilon_i}{k_B T}} \right) \right)$$

$I_A, I_B, I_C$ are the principal components of inertia tensor of the nonlinear molecule, $I$ is the degenerate moment of inertia for a linear molecule and $\sigma$ is the symmetry number of the molecule.

Dependence of the chemical potential on temperature and pressure can be obtained as the Gibbs free energy per molecule in the ideal gas approximation:

$$\mu(T,P) = G(T,P) = H(T) - TS(T,P) \qquad (S6)$$

For each gas considered in our work, we calculated the values of $\mu(T,P)$ according to the Eq. S6 and compared them with the experimental data from NIST JANAF databases (Supplementary Fig. 5). One can see that at given range of temperatures the precursors can be properly described with the presented approach.

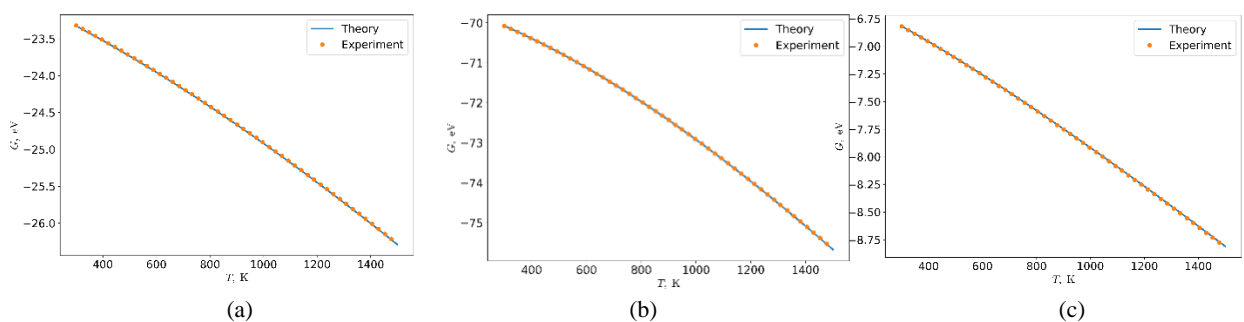

Supplementary Figure 5. Temperature dependence of Gibbs free energy of methane (a), borazine (b) and hydrogen per molecule. Experimental results from NIST JANAF thermochemical tables. Theoretical calculations were carried out using VASP and Atomic Simulation Environment. The reference energy values for experimental data are brought into compliance with DFT results.

## Thermochemical details: graphene growth

We calculated the values of chemical potential following the experimental CVD growth conditions of graphene monolayer[1–3] (Supplementary Table 1). Almost all the carbon chemical potential values derived from experimental growth conditions are much higher than the region of stability of monolayer graphene, despite the significant differences in the growth conditions in some of the presented experiments. This result indicates that the growth process is controlled by the kinetics of precursor decomposition and growth time. In other words, the thermodynamical equilibrium in the system is not achieved.

| Growth conditions | $\mu_C$, eV | Ref. |
|---|---|---|
| $CH_4 = 35$ sccm, $H_2 = 2$ sccm, pressure = 500 mTorr, temperature = 1000 °C | -7.41 | [64] |
| $p_{CH_4} = 10^{-3}$ Torr, $p_{H_2} = 0.1$ Torr, temperature = 1035 °C | -8.33 | [65] |
| $p_{CH_4} = 8$ mTorr, $p_{H_2} = 27$ mTorr, temperature = 1035 °C | -7.80 | [66] |
| $p_{CH_4} = 21$ mTorr, $p_{H_2} = 27$ mTorr, temperature = 1035 °C | -7.69 | [66] |
| Graphene monolayer | -9.59 to -9.34 | This work |
| Graphene quadlayer (bulk graphite) | > -9.34 | This work |

Supplementary Table 1. Comparison of the predicted stability regions of graphene and bulk graphite on Cu (111) substrate with the values of chemical potential derived from experimental growth conditions.

This fact becomes even more clear after Gibbs free energy of decomposition reaction is calculated:

$$\Delta G_r(P_{CH_4}, P_{H_2}, T) = 2G_{H_2(g)}(P_{H_2}, T) + F_{C(s)}(T) - G_{CH_4(g)}(P_{CH_4}, T) \quad (S7)$$

where $P_{CH_4}, P_{H_2}, T$ are partial pressures and temperature of the components. Here, we neglect a $pV$ term in Gibbs free energy of graphite, remaining only its Helmholtz energy. The result is represented in Supplementary Figure 6. First, we calibrate the computational methodologies on experimental data from NIST Thermochemical Tables at standard pressure (Supplementary Fig. 5a). Then we calculate the value of $\Delta G_r$ at $P_{CH_4} = 10^{-2}$ Torr, $P_{H_2} = 10^{-2}$ Torr, that are close to growth conditions (Supplementary Fig. 5b). The corresponding value of $\Delta G_r$ at $T = 1000\ °C$ is equal to -1.73 eV, which means that methane is thermodynamically unstable at growth conditions. In this way, if temperature and gas pressure remain constant, all the available methane decomposes into carbon layers and hydrogen molecules.

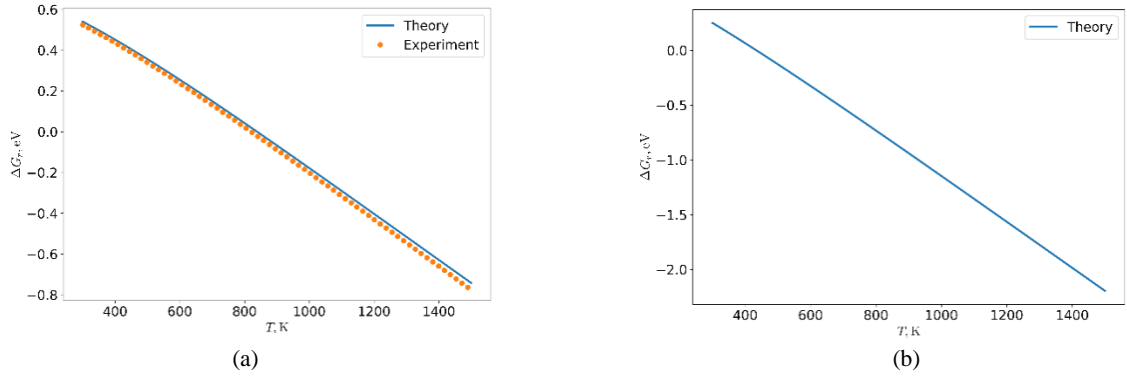

Supplementary Figure 6. Temperature dependence of Gibbs energy of reaction of methane decomposition at $P_{CH_4} = P_{H_2} = 1$ atm (a) and $P_{CH_4} = P_{H_2} = 10^{-2}$ Torr (b). Experimental values of $\Delta G_r$ are obtained from NIST JANAF database. Theoretical calculations were carried out using VASP and Atomic Simulation Environment. Our results show the values of $\Delta G_r < 0$ at graphene CVD growth conditions.

## Thermochemical details: h-BN growth

AB and borazine usage is notable because, since they do not exist in a gas phase at normal conditions and should be transported to the deposition chamber by a carrier gas (e.g. with a Ar/$H_2$ mixture with 5-25% of hydrogen). This fact makes the calculation of partial pressure of precursors quite non-trivial. The choice of ammonia borane makes the theoretical description even more sophisticated. Sublimation of AB leads to its decomposition into a wide range of boron- and nitrogen-containing molecules, including borazine and polyiminoborane[4–6] that serve as a source of atoms for h-BN growth. It is still debatable which particular molecule among the sublimation products is most likely to produce boron nitride. Each precursor's contribution depends on many parameters, such as the sublimation temperature, growth time, and substrate temperature[6]. Therefore, we assumed, for simplicity, that this role is played mostly by borazine.

If the growth of h-BN is performed with borazine as a precursor, the value of $P_{BZN}$ can usually be found in the literature. The situation is much more complicated if ammonia borane is chosen as a source of boron nitride, since no direct experimental data is presented on partial pressures neither of AB nor its decomposition products. The only data usually provided on AB is its sublimation temperature $T_{sub}$. Therefore, we estimated the value of $P_{BZN}$ as a saturated vapor pressure of ammonia borane at sublimation temperature. Then the value of $P_{BZN}$ can be calculated:

$$P_{BZN}(T_{sub}) \approx P_{AB}^{vap}(T_{sub}) = \exp\left(\frac{-\Delta_{sub}H_{298°}}{RT_{sub}} + \frac{\Delta_{sub}S_{298°}}{R}\right) \qquad (S8)$$

where the values of sublimation enthalpy and sublimation entropy are equal to $77.2 \pm 3.1$ kj mol$^{-1}$ and 142.56 J mol$^{-1}$ K$^{-1}$. Data on sublimation entropy and enthalpy is taken from work[5].

We derived the approximate values of $\mu_{BN}$ and CVD growth conditions and summarized them in Supplementary Table 2. As in the case of graphene, all the estimated values of chemical potential at experimental h-BN monolayer growth conditions turned to be much higher than those received from our calculation. It means that thermodynamically most stable structure at these conditions is bulk h-BN, and all available borazine in the system should decompose, forming an infinite number of h-BN layers, if only the conditions are kept constant. This result was actually expected since the Gibbs free energy of decomposition reaction given by Eq. S9 is < 0.

$$\Delta G_r(P_{BZN}, P_{H_2}, T) = 3G_{H_2(g)}(P_{H_2}, T) + 3F_{BN(s)}(T) - G_{BZN(g)}(P_{BZN}, T) \qquad (S9)$$

where $pV$ term in Gibbs free energy of bulk boron nitride is neglected. Temperature dependence of $\Delta G_r$ at standard pressure and at $P_{BZN} = 10^0$ Torr, $P_{H_2} = 10^2$ Torr that are close to experimental ones is given in Supplementary Figure 7. Following the same methodology as in case of graphene, we obtained the value of $\Delta G_r = -4.54$ eV at $T = 1000$ °C, which is quite consistent with our predictions.

| Precursor | Gas flow, sccm | Ar/H$_2$ ratio | P, Torr | $T_{gr}$,°C | $T_{sub}$,°C | $\mu_{BN}$, eV | Ref. |
|---|---|---|---|---|---|---|---|
| AB | 500 | 975:25 | 760 | 1070 | 90 | -17.14 | [73] |
| BZN | 13 (BZN), 2000 (H$_2$) | - | 760 | 1000 | - | -16.56 | [74] |
| AB | 75 | 100:25 | 760 | 1000 | 110 | -17.27 | [75] |
| AB | 200 | 85:15 | 760 | 1000 | 120 | -17.22 | [76] |
| AB | 100 | 90:10 | 760 | 1000 | 110 | -17.20 | [72] |
| AB | 100 | 975:25 | 760 | 1065 | 85 | -17.15 | [77] |
| h-BN monolayer | | | | | | -18.94 to -18.29 | This work |
| h-BN bilayer | | | | | | -18.29 to -18.27 | This work |

| | | | | | | > -18.27 | This work |
|---|---|---|---|---|---|---|---|
| h-BN four layers (bulk h-BN) | | | | | | | |

Supplementary Table 2. Comparison of the predicted stability regions of hexagonal boron nitride on Cu (111) substrate with the values of chemical potential derived from experimental growth conditions.

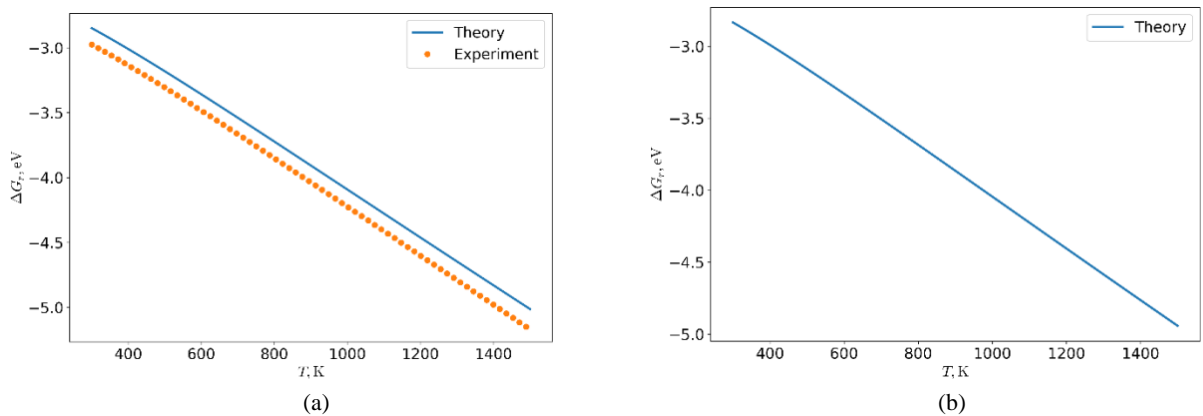

(a)    (b)

Supplementary Figure 7. Temperature dependence of Gibbs energy of reaction of borazine decomposition at $P_{BZN} = P_{H_2} = 1$ atm (a) and $P_{BZN} = 10^0$ Torr, $P_{H_2} = 10^2$ Torr (b). Experimental values of $\Delta G_r$ are obtained from NIST JANAF database. Theoretical calculations were carried out using VASP and Atomic Simulation Environment. Our results show the values of $\Delta G_r < 0$ at h-BN CVD growth conditions. Discrepancy between theory and experiment in (a) can be explained with insufficient accuracy of DFT predictions of $\Delta G_r$ at 0 K.